\documentclass[aps,prl,twocolumn,preprintnumbers]{revtex4}

\usepackage{amsmath,graphicx,color,amssymb,url}

\def\sgra{Sgr A${}^{*}$}

\def\beq{\begin{equation}}
\def\eeq{\end{equation}}

\newcommand{\nc}{\newcommand}
\nc{\Tr}{\mbox{Tr}}
\nc{\hc}{\mbox{H.c.}}
\nc{\Br}{\mbox{Br}}
\nc{\ev}{\;\mathrm{eV}}
\nc{\mev}{\;\mathrm{MeV}}
\nc{\gev}{\;\mathrm{GeV}}
\nc{\infinity}{\infty}

\nc{\JS}[1]{{\bf  \textcolor{blue}{[JS: {#1}]}}}
\nc{\SLS}[1]{{\bf  \textcolor{red}{[SLS: {#1}]}}}
\nc{\BDF}[1]{{\bf  \textcolor{green}{[BDF: {#1}]}}}

\begin{document}

\title{A black hole window into $p$-wave dark matter annihilation}

\author{Jessie Shelton} \author{Stuart L. Shapiro} \author{Brian D. Fields} 
\affiliation{Departments of Physics and of Astronomy,
  University of Illinois at Urbana-Champaign, Urbana, IL 61801, USA}

\begin{abstract}
  We present a new method to measure or constrain $p$-wave-suppressed
  cross sections for dark matter (DM) annihilations inside the steep
  density spikes induced by supermassive black holes.  We demonstrate
  that the high DM densities, together with the increased velocity
  dispersion, within such spikes combine to make thermal $p$-wave
  annihilation cross-sections potentially visible in gamma-ray
  observations of the Galactic center (GC).  The resulting DM signal
  is a bright central point source with emission originating from DM
  annihilations in the absence of a detectable spatially-extended
  signal from the halo.  We define two simple reference theories of DM
  with a thermal $p$-wave annihilation cross-section and establish new
  limits on the combined particle and astrophysical parameter space of
  these models, demonstrating that {\em Fermi} is currently sensitive
  to thermal $p$-wave DM over a wide range of possible scenarios for
  the DM distribution in the GC.

\end{abstract}

\maketitle

Most astrophysical searches for DM annihilation look for
velocity-independent, or $s$-wave, cross-sections $\langle\sigma
v\rangle$.  Theories with $p$-wave cross-sections, $\langle\sigma
v\rangle\propto v^2$, have largely remained out of reach in standard
searches for DM.  The only previous limits on thermal $p$-wave DM
annihilation come from cosmic microwave background (CMB) and radio
observations, which are sensitive to sub-GeV DM annihilating to
$e^+e^-$ \cite{Essig:2013goa}, while at higher masses CMB observations
are orders of magnitude away from sensitivity to thermal $p$-wave
annihilations \cite{Diamanti:2013bia}.

Here we show that thermal $p$-wave DM annihilation can be discovered
via gamma-ray emission within the density spikes that form around the
supermassive black holes (SMBHs) at the centers of DM halos.  The high
DM densities within such a BH-induced spike, and the increased
velocity dispersion required to support such densities, together boost
a $p$-wave-suppressed DM annihilation rate to potentially observable
values.  Such spikes would appear as point sources to gamma-ray
telescopes, and would contain sharp spectral features strongly
indicating a DM origin.  As a concrete illustration we consider the
Milky Way's SMBH \sgra, showing that {\em Fermi}'s observations of the
GC already place new constraints on $p$-wave DM annihilation, and
opening the door to potential discovery of thermal $p$-wave DM.

\medskip

\noindent {\bf {\em Models of thermal $p$-wave DM.}}
A thermal $p$-wave annihilation cross-section is a generic prediction
of a broad class of well-motivated DM models.  If DM is fermionic,
then in a parity-conserving theory its annihilation to spin-0,
$P$-even final states cannot receive any contribution from the
$s$-wave.  Thus, for instance, the well-known model of fermionic Higgs
portal DM \cite{Kim:2006af, Kim:2008pp} exhibits an annihilation
cross-section that receives its leading contribution in the $p$-wave.

We consider here two simple reference models of thermal $p$-wave dark
matter.  First is a {\em hidden sector Higgs portal} (HSHP)
model, where a Majorana DM particle $\chi$ annihilates to pairs of
dark scalars $s$.  We consider a minimal implementation of this model,
described by the Lagrangian $\mathcal{L} =\mathcal{L}_{kin}
-\frac{1}{2} y S \left(\chi \chi +\hc \right) + \frac{\mu_s ^ 2}{2}
S^2-\frac {\lambda_s} {4!} S ^ 4-\frac{\epsilon}{2}S ^ 2 |H | ^ 2$,
where a discrete symmetry $S\to -S$, $\chi\to i \chi$ forbids cubic
and linear terms in $V(S)$ and a Majorana mass term for $\chi$.  After
$S$ obtains a vacuum expectation value, it mixes with the Higgs,
allowing the mass eigenstate $s$ to decay to Standard Model (SM)
states.  The four free parameters describing this model can be taken
to be $m_\chi$, $m_s$, and the dimensionless Yukawa and portal
couplings $y$ and $\epsilon$.  For $m_s < 2m_h$, the branching ratios
of $s$ are given by the branching ratios of the SM Higgs at the same
mass, while for $m_s > 2 m_h \,(2 m_t)$ the decays $s\to h h\, (t\bar
t)$ must also be included.

Our second example is a {\em hidden sector axion portal} (HSAP) model,
where Majorana DM annihilates to pairs of pseudoscalars $a$, which can
be described by the simple Lagrangian $\mathcal {L} =\mathcal{L}_{kin}
- \frac{m_{\chi}}{2} \left(\chi \chi +\hc \right) -i\frac{1}{2} y a
\left(\chi \chi -\hc \right) - \frac{m_a^2}{2}a^2$; if an approximate
symmetry is responsible for protecting $m_a$, then higher-order
polynomial self-interactions of $a$ are suppressed \footnote{The
  higher-order process $\chi\chi\to 3a$ is $s$-wave, and while
  negligible at thermal freezeout, dominates DM annihilation today.
  We leave treatment of this mixed $s$+$p$-wave regime to future work
  (see also \cite{Amin:2007ir}), and focus on mass ranges where the
  three-body mode is kinematically forbidden.}.  The pseudoscalar can
decay via dimension-five couplings to SM gauge bosons. We will
consider the case where its coupling to gluons is absent at leading
order, which has the important consequence that $\Br(a\to
\gamma\gamma)$ is then $\mathcal{O}(1)$.  For simplicity, we will take
$\Br(a\to \gamma\gamma)=1$, neglecting (e.g.)  the generic but
model-dependent $\Br(a\to Z\gamma)$ that opens up for $m_a >
m_Z$. With this assumption, there are again four free parameters
describing this model, which can be taken to be $m_\chi$, $m_a$, $y$,
and $\epsilon\equiv m_a/\Lambda$, where $1/\Lambda$ is the
dimensionful coupling of the pseudoscalar to pairs of photons.

The range of gamma-ray signatures exhibited by these two theories
provides a representative guide to the observability of a
general model of thermal $p$-wave DM annihilating ultimately to
visible SM particles.  The thermally-averaged annihilation
cross-section in our two reference theories may be written
\begin{equation} 
\langle\sigma v\rangle =\frac {y^4v^2}{\pi m_\chi^2}
\sqrt{1-z}\, f(z) + \mathcal{O}(\sqrt{1-z}\, v^4),
\end{equation} 
where $v$ is the relative velocity, $z\equiv m_{s,a}^2/m_\chi^2$, and
$\sqrt{1-z} $ reflects final state phase space.  The function $f(z)$
is
\begin{equation} 
f_s(z) =  \frac{ 72-160 z+165 z^ 2-99 z^ 3+37 z^ 4-\frac{33}{4} z^{5}
    +\frac{27}{32} z^ {6}} {32
    (4-z)^2(2-z)^4} 
\end{equation} 
for the HSHP model, and
\begin{equation}
f_a(z) = \frac{(1-z)^2}{24 (2-z)^4} 
\end{equation} 
for the HSAP model.  We numerically solve the Boltzmann equation using
the full thermally-averaged expression for the cross-section to
determine the value of $y$ yielding the observed relic abundance,
leaving $m_\chi$ and $z$ as the free parameters of each model (we
assume for simplicity that the SM and the dark sector have the same
temperature at freezeout). This determines the thermal annihilation
cross-section $\sigma_{thermal}$ as a function of $m_\chi$ and
$z$. The coupling $\epsilon$ is irrelevant for astrophysical
signatures provided that (i) $\epsilon \ll y$ and (ii) the mediator
$(a,\,s)$ decays before BBN.

These models are both examples of secluded WIMPs
\cite{Pospelov:2007mp}; their direct detection and collider signals
are $\propto \epsilon^2$ and can be parametrically small.  The
indirect detection signal, on the other hand, remains directly tied to
the relic abundance, and thus for these models gamma rays from DM
annihilation in a BH spike can easily be the most robust route to
discovery.

\medskip

\noindent {\bf {\em The Milky Way's BH spike}}.
A DM density spike due to the presence of the central SMBH \sgra\
forms inside the radius of gravitational influence of the SMBH, $r_h =
M/v_0^2$ ($G\equiv 1$). Here $M$ is the mass of the SMBH and $v_0$ is
the velocity dispersion of DM in the halo outside the spike.  The
precise form this spike takes depends on both the properties of DM and
the formation history of the BH, making DM annihilation
signals from such spikes potentially powerful probes of both the
particle properties of DM and the evolution of the Milky Way.

A general DM spike $\rho(r)$ may be well-approximated by a series of
connected power-law profiles.  The spike begins its growth from the
inner halo, which in a generalized NFW halo follows a power law
$\rho(r)=\rho(r_0) (r_0/r)^{\gamma_c}$.  DM-only simulations yield
typical values of $0.9\lesssim \gamma_c\lesssim 1.2$
\cite{Diemand:2008in, Navarro:2008kc}, while the dissipative collapse
of baryons into the disk can adiabatically contract the central DM
halo into a steeper power-law \cite{Blumenthal:1985qy, Gnedin:2004cx,
  Gustafsson:2006gr}, with values as high as $\gamma_c \sim 1.6$ being
possible in the Milky Way \cite{Pato:2015dua}.

The DM spike grows inside $r_b \approx 0.2 r_h$ \cite{Gondolo:1999ef},
and is well-described by $\rho_{sp}(r) = \rho(r_{b})
(r_b/r)^{\gamma_{sp}}$.  
The power-law
index $\gamma_{sp}$ depends on possible formation histories. The
steepest spikes are formed by the response of a collisionless DM halo
to the adiabatic growth of a central SMBH, in which case $\gamma_{sp}
= (9-2\gamma_c)/(4-\gamma_c)$, which for $0 < \gamma_c \leq 2$ yields
$2.25 < \gamma_{sp}< 2.5$ \cite{Pee72a,Gondolo:1999ef,Mer04,GneP04}.
The heating of DM from gravitational scattering off of a sufficiently
dense and cuspy stellar density within $r_h$ could substantially
soften the DM spike over the lifetime of the Milky
Way~\cite{Mer04,GneP04}.  The final equilibrium spike profile attained
as a result of this stellar heating has $\gamma_{sp} = 1.5$, while a
spike that is in the process of being heated would have an
intermediate value of $\gamma_{sp}$, perhaps $\gamma_{sp}\sim 1.8$;
see \cite{Fields:2014pia} for a recent summary and discussion,
including other possible spike and halo solutions.

Once the DM spike attains the ``annihilation plateau'' density
$\rho_{ann} = m_{\chi}/\langle\sigma v\rangle t$ at 
$r\equiv r_{in} $, DM annihilations become relevant
over the lifetime $t$ of the spike ($\approx$ the age of the SMBH).
For $r< r_{in}$, annihilations weaken the power law growth to
$\rho_{in}(r) = \rho_{ann} (r_{in}/r)^{1/2}$ \cite{Vasiliev:2007vh}.
In this innermost region only particles in eccentric orbits with
apocenters outside $r_{in}$ contribute significantly to the density
inside $r_{in}$.  This weakened profile arises whenever the timescale
for annihilation in the Galaxy lifetime decreases with decreasing $r$
in a canonical spike, which is the case for $p$-wave as well as
$s$-wave annihilations.  Finally, the inner boundary of the spike is
at $4M$ \cite{Sadeghian:2013laa, Shapiro:2014oha}.
The resulting density profile thus may be written as
\cite{Fields:2014pia}
\begin{eqnarray}
\rho(r) &=& 0, \ \ \ r < 4M \ \ ({\rm capture \ region}), \\
&=& \frac{\rho_{\rm sp}(r)\rho_{\rm in}(t,r)}
{\rho_{\rm sp}(r) + \rho_{\rm in}(t,r)}, 
\ 4M < r < r_b \ \ ({\rm spike}), \nonumber \\
&=& \rho_b(r_b/r)^{\gamma_{c}}, \ \ \ \ \ \ r_b < r < R_H \ \ \
({\rm cusp}), \nonumber \\
&=& \rho_H (R_H/r)^{\gamma_H}, \ \ \ R_H < r \ \ \ 
({\rm outer \ halo}), \nonumber
\end{eqnarray}
The dominant contribution to any annihilation signal comes from the
region where $r\sim r_{in}$.  

Critically, the velocity dispersion increases inside the spike to
support the power-law increase in density.  For the (isotropic)
velocity dispersion profile, we match an approximate, piece-wise
continuous solution of the Jeans equation in the spike onto a constant
dispersion in the inner halo,
\begin{eqnarray}
v^2(r) &=&  \frac{M}{r} \frac{1}{1+\gamma_{in}} 
\left[
1 + \frac{r}{r_{\rm in}}
\left( \frac{\gamma_{in}-\gamma_{sp}}{1+\gamma_{sp}} \right)
\right], \nonumber \\
&& \ \ \ \ \ \ \ \ \ \ \ \ \ \ \ \ 4M \leq r < r_{\rm in} \ \
({\rm inner \  spike}), \label{eq:velocity}\\
&=& \frac{M}{r} \frac{1}{1+\gamma_{sp}},  \ \ \ 
r_{\rm in} \leq r < \frac{r_h}{1+\gamma_{sp}} \ \ ({\rm outer \  spike}), 
 \nonumber \\
&=& v_0^2 = {\rm const},   \ \ \frac{r_h}{1+\gamma_{sp}}\leq r 
\ \ ({\rm cusp \ \& \ outer \ halo}). \nonumber
\end{eqnarray} 
Following \cite{Fields:2014pia}, we adopt the following parameter
values for the Milky Way's DM halo and SMBH:
$M = 4 \times 10^6 {\rm M_{\odot}}$~\cite{GenEG10,Ghez.etal08},
$\rho_\odot 
= 0.3 \pm 0.1{\rm ~GeV ~cm^{-3}}$~\cite{BovT12}, $v_0 = 105 \pm 20
{\rm ~km~s^{-1}}$~\cite{Gultekin:2009qn}, $R_{\odot} =
8.46^{+0.42}_{-0.38}$ kpc~\cite{Do:2013upa}, and $t_{\rm ann} =
10^{10}$ yrs. With these parameters we find a spike radius of $r_h =
1.7$ pc, subtending $0.012^{\circ}$, well below the resolution of
current and future gamma-ray telescopes~\cite{fermiangular, agile,
  magic, veritas, hess, cta, hawc, gammalight, dampev1}.    
The remaining parameters of the spike solution are the exponents
$\gamma_c$, $\gamma_{sp}$; with no direct measurements of these
quantities, we treat them as free parameters of our model.
Typical values for $r_{in}$ fall in the range $10^{-3}$--$10^{-5}$ pc.

\medskip

\begin{figure}
\begin{center}
\includegraphics[width=0.9\linewidth]{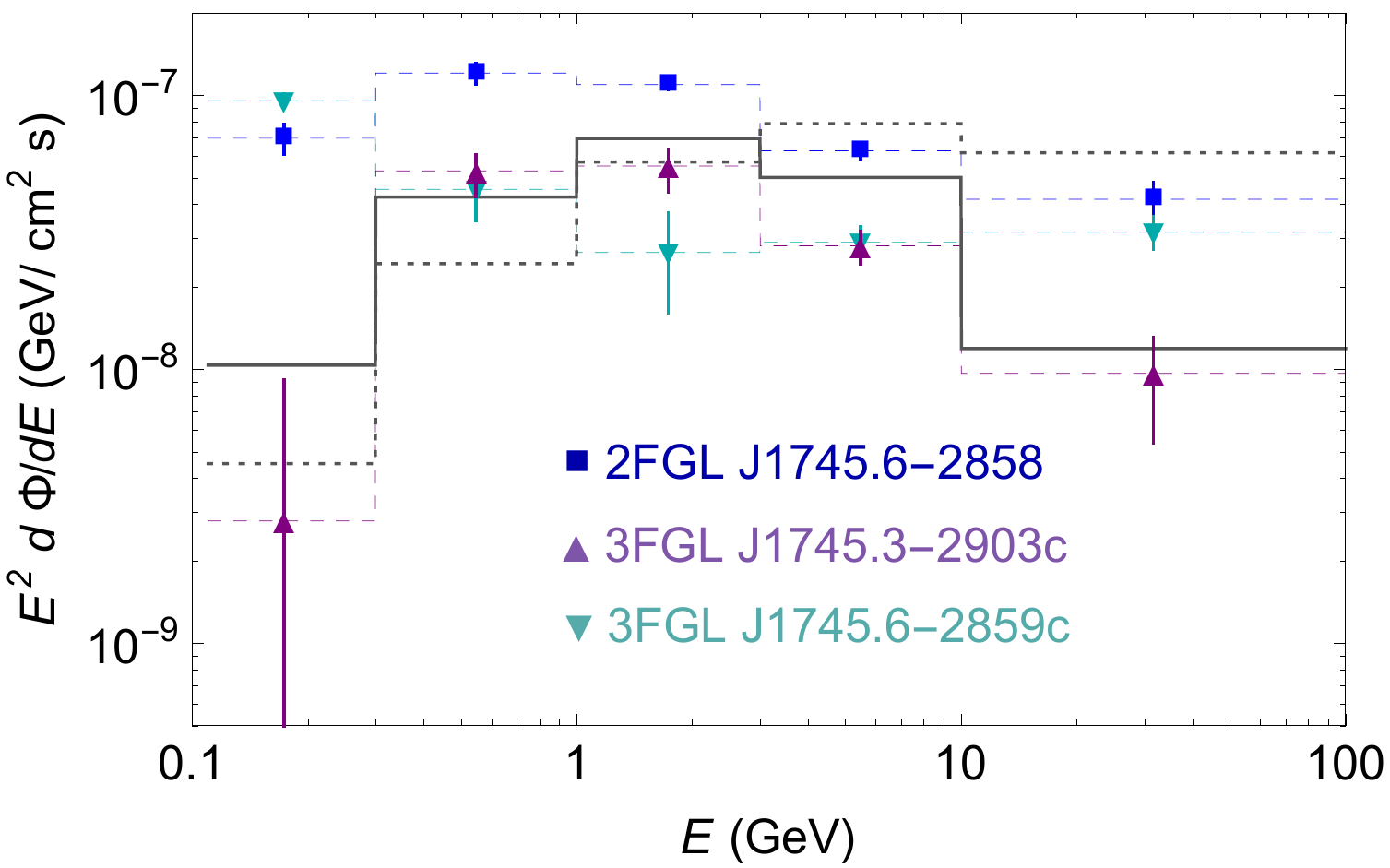}
\caption{Spectra of three possible candidates to contain emission from a BH-induced DM
  density spike
  \cite{Fermi-LAT:2011iqa, 3FGL}.  In black are two example
  predictions of thermal HSHP DM with different particle
  masses (determining the shape) and spike and
  halo parameters (controlling the normalization).  The
  solid line shows predictions for $m_\chi = 45$ GeV, $m_s = 10$ GeV,
  $\gamma_c = 1.3$, and $\gamma_{sp}=1.8$; the dashed line, $m_\chi =
  110$ GeV, $m_s = 50$ GeV, and an adiabatic spike in a halo with
  $\gamma_c = 1.1$. 
  \label{fig:pointsourcecomparison}}
\end{center}
\end{figure}

\noindent {\bf {\em Observability of $p$-wave DM: continuum emission.}}  
Our aim here is to demonstrate that there
is a sizable range of possible spike and halo parameters for which
emission from thermal HSHP DM annihilation in a BH spike at the GC is
comparable or greater in brightness to detected gamma-ray point
sources in the same region.
{\em Fermi}'s Third Point Source Catalogue (3FGL) \cite{3FGL} contains
several point sources near the GC.  The {\em Fermi-LAT} localization
accuracy for a point source depends on its brightness and that of the
surrounding diffuse emission, and is $\sim 9 \ \rm arcmin$ for the
sources in question.  The {\em Fermi} team has associated source
\texttt{3FGL J1745.6-2859c} with \sgra.  This has an integrated flux
of $ 2.18\times 10 ^{-8}\mathrm{photons}/\mathrm{cm}^2\mathrm{s}$ in
the energy range 1-100 GeV.  The source \texttt{3FGL J1745.3-2903c} is
slightly brighter, with an integrated flux $\Phi = 3.87\times 10 ^
{-8}\, \mathrm{photons}/\mathrm{cm}^2\mathrm{s}$ in the same energy
range, but is 5.1 arcsec offset from \sgra, and so is a less likely
association.  Spectra of these two sources appear in
Fig.~\ref{fig:pointsourcecomparison}, together with spectra from HSHP
DM annihilating inside possible BH spikes.  The remaining bright 3FGL
source within 30 arcmin of \sgra\ is associated with a pulsar wind
nebula and thus is not a BH spike candidate.
%
%
%
{\em Fermi}'s Second Point Source Catalogue \cite{Fermi-LAT:2011iqa}
reported a single central point source \texttt{2FGL J1745.6-2858},
previously identified as a BH spike candidate \cite{Fields:2014pia},
and shown for comparison.  

\begin{figure}
\begin{center}
\includegraphics[width=\linewidth]{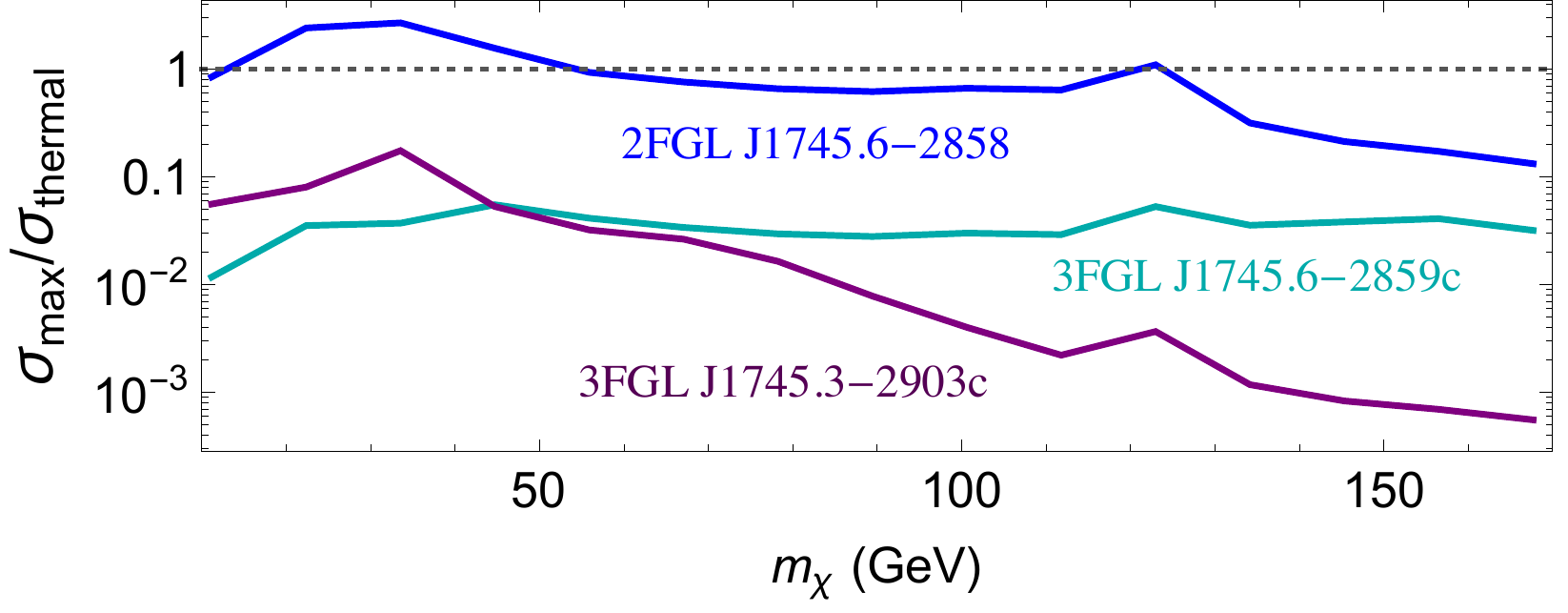}
\includegraphics[width=\linewidth]{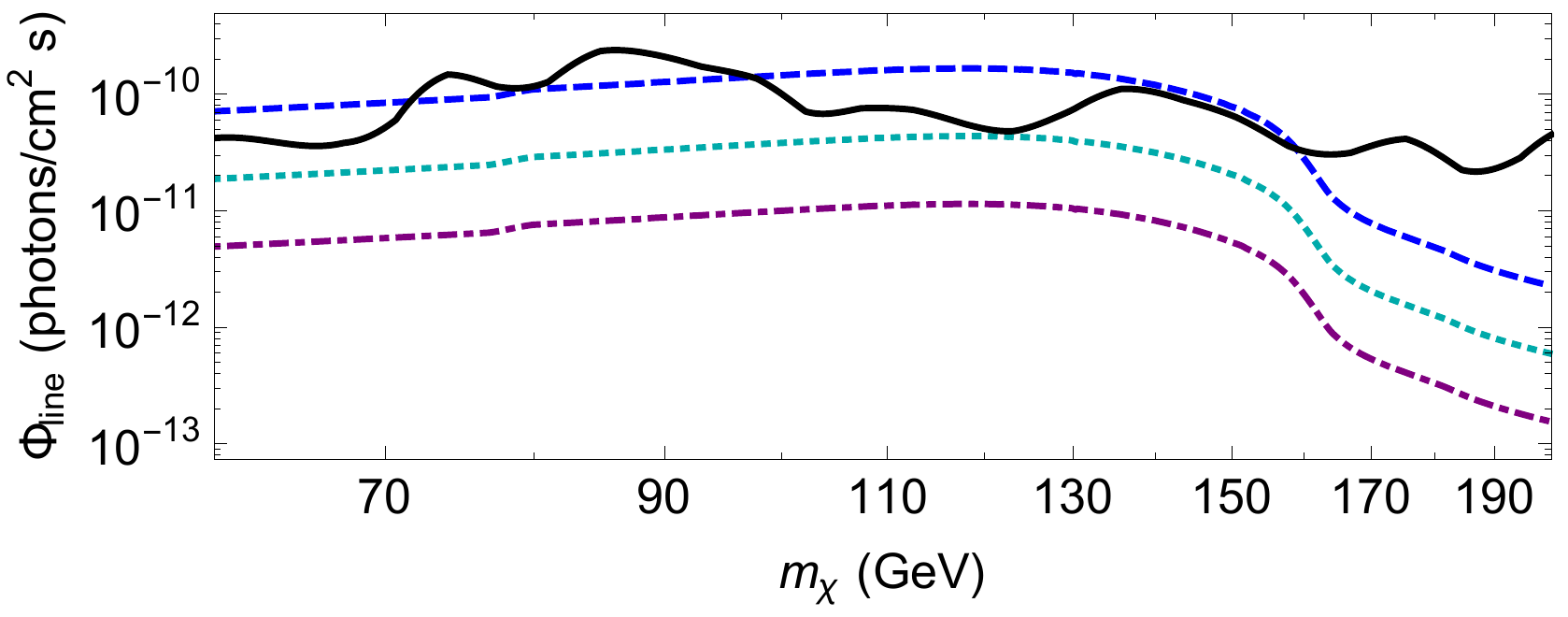}
\caption{Sensitivity to HSHP DM from Fermi observations of the
  GC. Top: the minimum value of the DM annihilation cross-section for which
  the primary photon emission exceeds observed point source spectra at
  $95\%$ CL in at least one bin, as a function of DM mass,
  for adiabatic spikes with $\gamma_c=1.1$ and $z=0.2$.  Bottom: the
  predicted flux in a box-shaped spectral feature from HSHP DM
  annihilating inside adiabatic spikes for $\gamma_c = $ 1.3 (blue, dashed),
  1.2 (cyan, dotted), 1.1 (purple, dash-dotted).  Results are shown for $z=0.99$, where the
  box is sufficiently narrow to appear as a line, and compared to {\em
    Fermi}'s limit \cite{Ackermann:2013uma} on line flux from the GC
  (black). 
  \label{fig:HSHP}}
\end{center}
\end{figure}

\begin{figure}
\begin{center}
\includegraphics[width=0.98\linewidth]{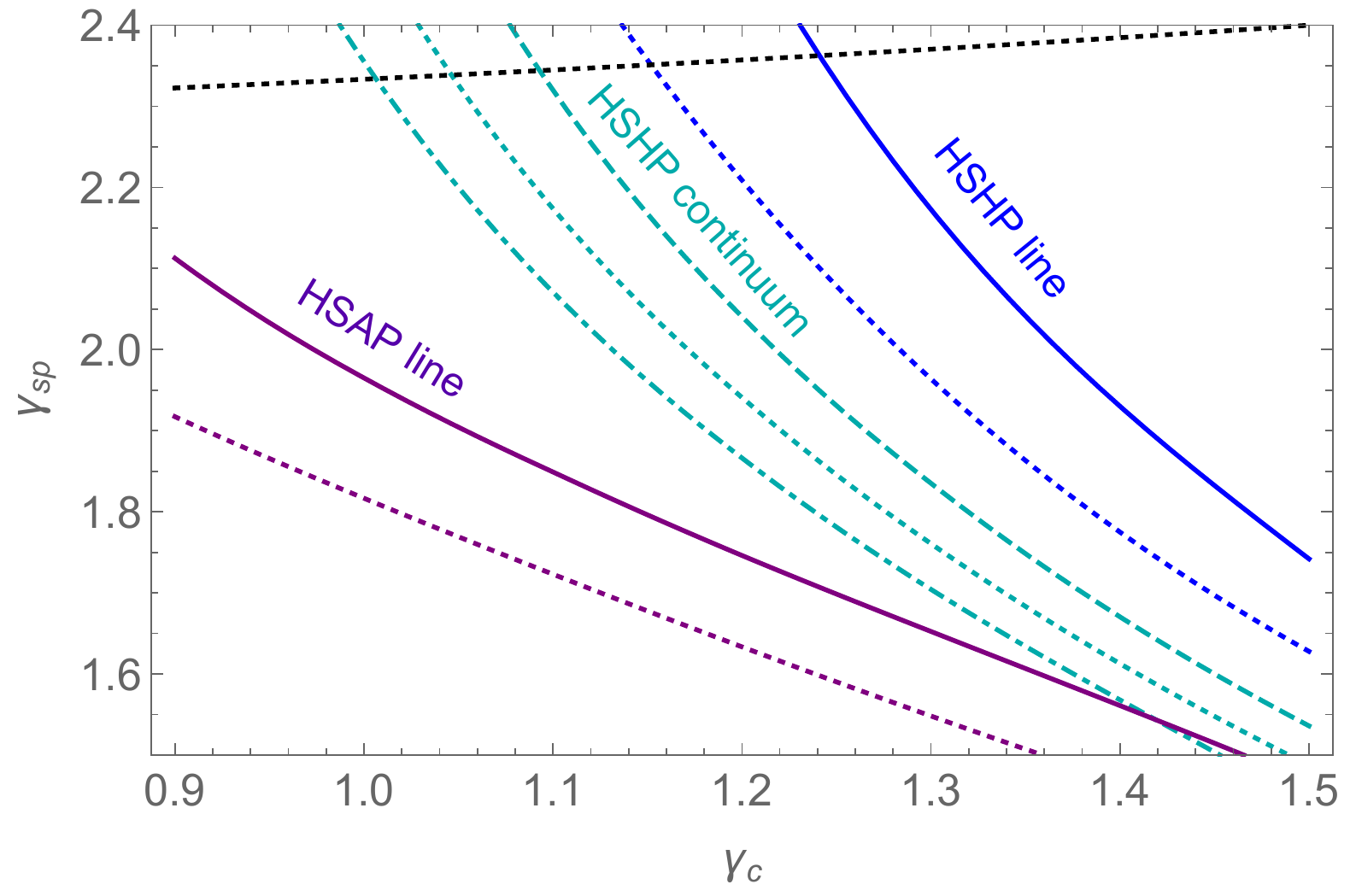}
\caption{The region of combined halo$+$spike parameter space where
  {\em Fermi} observations constrain thermal $p$-wave annihilation
  cross-sections for $m_\chi = 110$ GeV.  In cyan are
  continuum constraints for HSHP DM ($z=0.2$), from comparison with 2FGL J1745.6-2858, 3FGL J1745.6-2959c, and 3FGL J1745.3-2903.c in the dashed, dotted, and dot-dashed lines respectively; 
  the blue solid line shows the line search limit of Ref.~\cite{Ackermann:2013uma}
  on HSHP DM ($z=0.99$); the purple solid line shows the line
  search limit on HSAP DM ($z=0.99$).  Dotted blue and purple lines
  show an estimate of the improved sensitivity to HSHP and HSAP DM respectively offered by a
  dedicated line search. Regions
  above the curves exceed observations.  Adiabatic spikes are
  indicated by the dotted black line. 
  \label{fig:messageplot}}
\end{center}
\end{figure}

In Figs. 2 and 3 we compare the flux from HSHP DM to the point source
fluxes detected by {\it Fermi}.  For each of the three point sources, we
find the minimum value of the cross-section such that the primary
photon flux from DM annihilations exceeds the observed flux in any
energy bin at more than $95\%$ CL, treating each bin as an independent
Poisson-distributed variable.  We use \texttt{Pythia 8}
\cite{Sjostrand:2007gs} to generate photon spectra. Scalar branching
ratios in the range $m_s < 75$ GeV are calculated using
\texttt{HDECAY} \cite{Djouadi:1997yw}; for $m_s > 75$ GeV, we use
branching ratios from the LHC Higgs Cross-Section Working Group
\cite{Heinemeyer:2013tqa}.  Our results are shown in the top panel of
Fig.~\ref{fig:HSHP} for the representative choice of mass-squared
ratio $z=0.2$.  The large differences between the maximum
cross-section allowed by the three different point sources arise
because the flux from a BH spike is only weakly dependent on the
annihilation cross-section; for a $p$-wave spike, $\Phi \propto
(\sigma/\sigma_{\mathrm{thermal}})^{(3-\gamma_{sp})/(1+\gamma_{sp})}$.
This exponent is $\sim 1/5$ for adiabatic spikes, and $\sim 1/2$ for
heated spikes. The BH spike would outshine detected point sources
across a wide range of possible spike and halo scenarios, notably
including adiabatic spikes for $\gamma_c\gtrsim 1$, as we show in
Fig.~\ref{fig:messageplot}.

Greater sensitivity could be obtained in a dedicated search using more
sophisticated signal and background modeling.  Signal will also
include sizeable secondary emission arising from the interaction of DM
annihilation products with ambient dust, starlight, and magnetic
fields.  Meanwhile, a given BH spike candidate will generically
contain astrophysical emission in addition to any DM signal.  The 3FGL
`variability index' for \texttt{3FGL J1745.6-2859c} suggests that it
may be time-variable, which, if confirmed, would set a floor for the
astrophysical contribution to the gamma-ray flux from this source.

\medskip

\noindent {\bf {\em Limits on $p$-wave DM:  box and line searches.}}
Given the large systematic uncertainties on DM halo and spike
distributions, it is difficult to conclusively discover or exclude
$p$-wave DM using a continuum signal alone.  Even for DM with an
$s$-wave annihilation cross-section, where support for a DM
interpretation of a potential gamma-ray signal may be obtained from
its extended spatial distribution, the subdominant but sharp gamma-ray
line at $E_\gamma = m_{DM}$ remains a smoking gun for a DM origin,
in contrast to the broad continuum
signature which may more easily be mimicked by astrophysical processes
\cite{Bergstrom:1988fp,Bergstrom:1997fj}.  The analog of a gamma-ray
line in our reference models is a gamma-ray {\em box}, from the decay
of a (boosted) mediator to a pair of photons \cite{Ibarra:2012dw}.
The upper and lower endpoints of the box depend on the mass splitting
between the DM and the mediator, and are given by $ E_\gamma^{\pm} =
\frac{m_\chi} {2}\left(1\pm\sqrt{1-z}\right).  $ In the
near-degenerate limit $z\approx 1$, the box becomes narrower than {\it
  Fermi}'s energy resolution, and limits from line searches may be
directly applied.  Sensitivity to the flux in wider boxes in less
degenerate spectra is $\sim 2$-$5$ times weaker than the sensitivity
to the flux in line-like features at the same value of $m_\chi$
\cite{Ibarra:2012dw,Ibarra:2015tya}.
%

%
The line search with the best sensitivity to $p$-wave DM is region R3
from \cite{Ackermann:2013uma}, which considers the inner 3$^\circ$
around the GC and notably does not mask point sources. We reinterpret
this search as a constraint on narrow boxes ($z=0.99$) originating
from a BH spike \footnote {We have checked that the broadening of the
  box from increased velocity dispersions within the spike is
  negligible given the energy resolution of realistic gamma-ray
  telescopes.}.  We show the resulting exclusions for HSHP DM
annihilating in adiabatic spikes in the bottom panel of
Fig.~\ref{fig:HSHP}. In this model, $\Br(s\to \gamma \gamma) \lesssim
10^{-3}$, making the current line search less sensitive than the
continuum constraints; this conclusion would also apply to line
searches in fermionic Higgs portal models, and to HSAP models where
the pseudoscalar dominantly decays to gluons.  Thus a DM origin for a
potential Milky Way signal in Higgs portal models would be established
via the discovery of a sharp spectral feature within the emission of a
previously-discovered point source.  For our HSAP model, however, the
box is the leading signal, resulting in much greater sensitivity.  Our
limits on HSAP DM are shown in Fig.~\ref{fig:messageplot}.  In this
model the non-optimized line search of Ref.~\cite{Ackermann:2013uma}
is already sufficiently powerful to exclude adiabatic spikes given the
Galactic parameters adopted here.

As a rough estimate of the potential improvement offered by a
dedicated search for lines near \sgra\, we approximate the gain in
significance as $\sqrt{B_{3}/B_{0.3}}$, where $B_{0.3}$ is the
background flux in a search region of radius of order the angular
resolution for high-energy photons, $0.3^\circ$, and $B_{3}$ is the
background flux in {\em Fermi}'s search region R3.  Examining SOURCE
class photons with energies above 10 GeV gives an estimate for this
ratio of $B_3/B_{0.3} \approx 10$.
We show the resulting estimate of potential sensitivity to thermal
$p$-wave DM annihilation in the dotted lines in
Fig.~\ref{fig:messageplot}.

\medskip

\noindent {\bf\em{Summary and Conclusions.}} 
High densities in DM spikes around SMBHs, together with the enhanced
velocity dispersions required to support them, allow
$p$-wave-suppressed DM annihilation cross-sections to yield visible
signals in current gamma-ray telescopes. Using {\em Fermi}'s
observations of the GC, we placed entirely novel constraints on
thermal $p$-wave annihilation cross-sections.  More precisely, we
constrained a sizeable range of combined particle and astrophysical
models (in much the same spirit as Galactic searches for $s$-wave
annihilation that depend on a halo model as well as the final state),
and established a well-motivated range of particle and astrophysical
parameter space where DM discovery may be uniquely possible via the
detection of sharp spectral features in a central gamma-ray point
source.

Beyond the Milky Way, most bulge galaxies are expected to host
SMBHs, which will in turn create DM density spikes of varying
  steepness.  DM annihilation within these spikes yields gamma-ray
point sources with a common spectrum of primary photons, although the
secondary emission will depend on the local environment of each SMBH.
This conclusion is true for $s$-wave as well as $p$-wave DM; the
novelty for $p$-wave DM is that the point sources may be observable
{\em even in the absence of a detectable halo signature}.  When the
point sources are too dim to be resolved, BH spikes provide a novel
mechanism for $p$-wave DM to contribute at potentially nonnegligible
levels to the extragalactic diffuse background.

\medskip

\noindent {\it Acknowledgments}.  We are happy to acknowledge useful
conversations with J.~Evans and S.~Ritz and are grateful to
D.~Thompson for insight into the {\em LAT} astrometry.  We
particularly thank T.~Linden for informing our estimate of the
improvement available in an optimized line search.  This work was
supported in part by NSF Grant PHY-1300903 as well as NASA Grant
NNX13AH44G at the University of Illinois at Urbana-Champaign. JS is
grateful to the Mainz Institute for Theoretical Physics for its
hospitality and partial support during the completion of this work.

\bibliography{spikebib-p}

\end{document}